\PassOptionsToPackage{hyphens}{url}
\PassOptionsToPackage{colorlinks=false,breaklinks=true}{hyperref}

\documentclass[prd,twocolumn,superscriptaddress,floatfix,nofootinbib]{revtex4-2}

\usepackage[T1]{fontenc} % if needed
\usepackage{amsmath,amssymb,amsfonts,amsthm}
\usepackage{graphicx}
\usepackage[usenames,dvipsnames]{color}
\usepackage{enumitem}
\usepackage{verbatim}
\usepackage[percent]{overpic}
\usepackage{rotating}
\usepackage{array}
\usepackage{mathtools}
\usepackage{lmodern}
\usepackage[normalem]{ulem}
\usepackage{braket}
\usepackage{tensor}
\usepackage{dsfont}
\usepackage{bm}
\usepackage{tikz}
\usepackage[version=3]{mhchem}
\usepackage[utf8]{inputenc}

\newcommand{\simgt}{\lower.5ex\hbox{$\; \buildrel > \over \sim \;$}}
\newcommand{\simlt}{\lower.5ex\hbox{$\; \buildrel < \over \sim \;$}}
\usepackage{mathrsfs}
\usetikzlibrary{decorations.pathmorphing}
\usepackage{CJKutf8}



\usepackage{hyperref}
\hypersetup{colorlinks=false,breaklinks=true}

\begin{document} 

\title{Space-based cm/kg-scale Laser Interferometer for Quantum Gravity}
\author{ Nobuyuki Matsumoto} \thanks{Corresponding author: \texttt{matsumoto.granite@gmail.com}}
\affiliation{\small\it Department of Physics, Faculty of Science, Gakushuin University, 1-5-1, Mejiro, Toshima, Tokyo, 171-8588 Japan}

\author{~~Katsuta Sakai}
\affiliation{\small\it Department of Physics, Faculty of Science, Gakushuin University, 1-5-1, Mejiro, Toshima, Tokyo, 171-8588 Japan}
\affiliation{\small\it TRIP Headquarters, RIKEN, Wako 351-0198, Japan}

\author{Kosei Hatakeyama}
\affiliation{\small\it Department of Physics,  Kyushu University, 744 Motooka, Nishi-Ku, Fukuoka 819-0395 Japan}

\author{ Kiwamu Izumi}
\affiliation{\small\it Institute of Space and Astronautical Science, Japan, Aerospace Exploration Agency, 3-1-1 Yoshinodai, Chuo, Sagamihara, Kanagawa 252-5210 Japan}

\author{Daisuke Miki}
\affiliation{\small\it The Division of Physics, Mathematics and Astronomy, California Institute of Technology, Pasadena, CA 91125, USA}

\author{~~Satoshi Iso} 
\affiliation{RIKEN Center for Interdisciplinary Theoretical and Mathematical Sciences (iTHEMS), RIKEN, Wako 351-0198, Japan
}
\affiliation{\small\it Theory Center, High Energy Accelerator Research Organization (KEK), Oho 1-1, Tsukuba, Ibaraki 305-0801 Japan }
\affiliation{\small\it Graduate University for Advanced Studies (SOKENDAI), Oho 1-1, Tsukuba, Ibaraki 305-0801 Japan}
%\affiliation{\small\it International Center for Quantum-field Measurement Systems for Studies of the Universe and Particles (QUP), KEK, Oho 1-1, Tsukuba, Ibaraki 305-0801 Japan}

\author{Akira Matsumura}
\affiliation{\small\it Department of Physics,  Kyushu University, 744 Motooka, Nishi-Ku, Fukuoka 819-0395 Japan}
\affiliation{\small\it Quantum and Spacetime Research Institute, Kyushu University, 744 Motooka, Nishi-Ku, Fukuoka 819-0395 Japan}

\author{ Kazuhiro Yamamoto}
\affiliation{\small\it Department of Physics,  Kyushu University, 744 Motooka, Nishi-Ku, Fukuoka 819-0395 Japan}
\affiliation{\small\it Quantum and Spacetime Research Institute, Kyushu University, 744 Motooka, Nishi-Ku, Fukuoka 819-0395 Japan}

%\affil[6]{\small\it Department of Physics,  Kyushu University, 744 Motooka, Nishi-Ku, Fukuoka 819-0395 Japan}
%\affil[7]{\small\it College of Liberal Arts and Sciences, Tokyo Medical and Dental University,2-8-30 Kounodai, Ichikawa, Chiba 272-0827, Japan}
%\affil[8]{\small\it Research Center for Advanced Particle Physics, Kyushu University, 744 Motooka, Nishi-ku, Fukuoka 819-0395 Japan}

\begin{abstract}
The experimental verification of the quantum nature of gravity represents a milestone in quantum gravity research. Recently, interest has grown for testing it via gravitationally induced entanglement (GIE). Here, we propose a space-based interferometer inspired by the LISA Pathfinder (LPF). Our design employs two kg-scale gold-platinum test masses which, unlike in the LPF, are surrounded by a shield below 1 K and positioned side-by-side with a centimeter-scale separation. This configuration enables the detection of GIE through simultaneous measurements of differential and common-mode motions.
To estimate the integration time required for GIE detection, we simulate quantum measurements of these modes, considering noise sources such as gas damping, black-body radiation, and cosmic-ray collisions. 
Our results show that GIE can be demonstrated with a few modifications to the LPF setup. 
\end{abstract}

\maketitle
\flushbottom

{\it Introduction---}
Although extensive research on quantum gravity has been conducted, limited evidence has been obtained from astronomical observations~\cite{PhysRevLett.116.221101,Berti_2015,PhysRevLett.109.241104,PhysRevD.87.122001,PhysRevLett.133.071501,PhysRevLett.83.2108,PhysRevD.99.083009}. This is primarily because the Planck scale, where quantum-gravity effects become significant, is far beyond the scales achievable in current experiments. Recently, novel methods have been proposed to bypass this difficulty and test whether Newtonian gravity exhibits quantum properties in the non-relativistic regime~\cite{PhysRevLett.119.240401,Marletto:2017kzi}. These methods are based on a fundamental theorem in quantum information theory, which states that quantum entanglement cannot be generated through the LOCC (Local Operations and Classical Communication)~\cite{RevModPhys.81.865}. Demonstrating that entanglement can be generated via gravity would provide direct evidence of its quantum nature.

Toward this end, several experiments have been proposed~\cite{PhysRevA.102.062807, PhysRevResearch.3.023178,Qvarfort_2020,PhysRevA.98.043811,PRXQuantum.2.030330,PhysRevLett.128.110401,PhysRevD.110.024057,Krisnanda2020,PhysRevD.108.106014,PhysRevResearch.6.013199,Miao,Datta}; however, due to the weak nature of gravity, achieving this remains a challenge. There are three possible approaches to overcome this challenge: (1) signal amplification~\cite{PhysRevD.108.106014,PhysRevLett.128.110401,Krisnanda2020}, (2) direct noise reduction~\cite{PhysRevA.102.062807,PhysRevResearch.3.023178,Qvarfort_2020,PhysRevA.98.043811,PhysRevResearch.6.013199,PhysRevD.110.024057, PhysRevLett.122.071101}, and (3) development of noise-tolerant measurement methods~\cite{PRXQuantum.2.030330,Miao,Datta,PhysRevResearch.3.023178}. The experimental setups can be categorized into two types: one targets massive objects, where gravitational interactions can dominate more easily, particularly in optomechanical systems~\cite{PhysRevD.108.106014,Miao,PhysRevA.98.043811,PhysRevD.110.024057, PhysRevLett.122.071101}, while the other focuses on microscopic systems~\cite{PhysRevA.102.062807, PhysRevResearch.3.023178,Qvarfort_2020,PhysRevResearch.6.013199,PhysRevResearch.3.023178}, where quantum control is more easily achievable. A system combining these two components is also proposed~\cite{PhysRevLett.128.110401,PRXQuantum.2.030330}. 

Here, we propose a cavity-free laser interferometer that optimizes both signal amplification and direct noise reduction in optomechanical systems, in contrast to previous cavity-based studies~\cite{PhysRevD.108.106014,Miao,PhysRevA.98.043811,PhysRevD.110.024057,PhysRevLett.122.071101}. 
Our proposed experiment closely resembles the space-based interferometer LISA Pathfinder (LPF)~\cite{PhysRevLett.116.231101}, which
marked the beginning of a new era in interferometric experiments; however, it differs in the following three key
aspects: (1) The test masses are placed close to enable the generation of gravitationally induced entanglement (GIE);
(2) Both differential and common-mode motions are measured to enable the detection of GIE via Kalman filtering; and
(3) The masses are surrounded by a shield below 1 K, and both modes of motion are further cooled via feedback using a high-pass filter. 
Using the space environment, the resonance frequency of the test mass can be reduced to 0.25 mHz, effectively amplifying the gravitational coupling signal while remaining free from ground vibrations, thereby making it possible to observe the GIE within a total integration time of about 40 days.

\begin{figure}
    \centering
    \includegraphics[width=8.5cm]{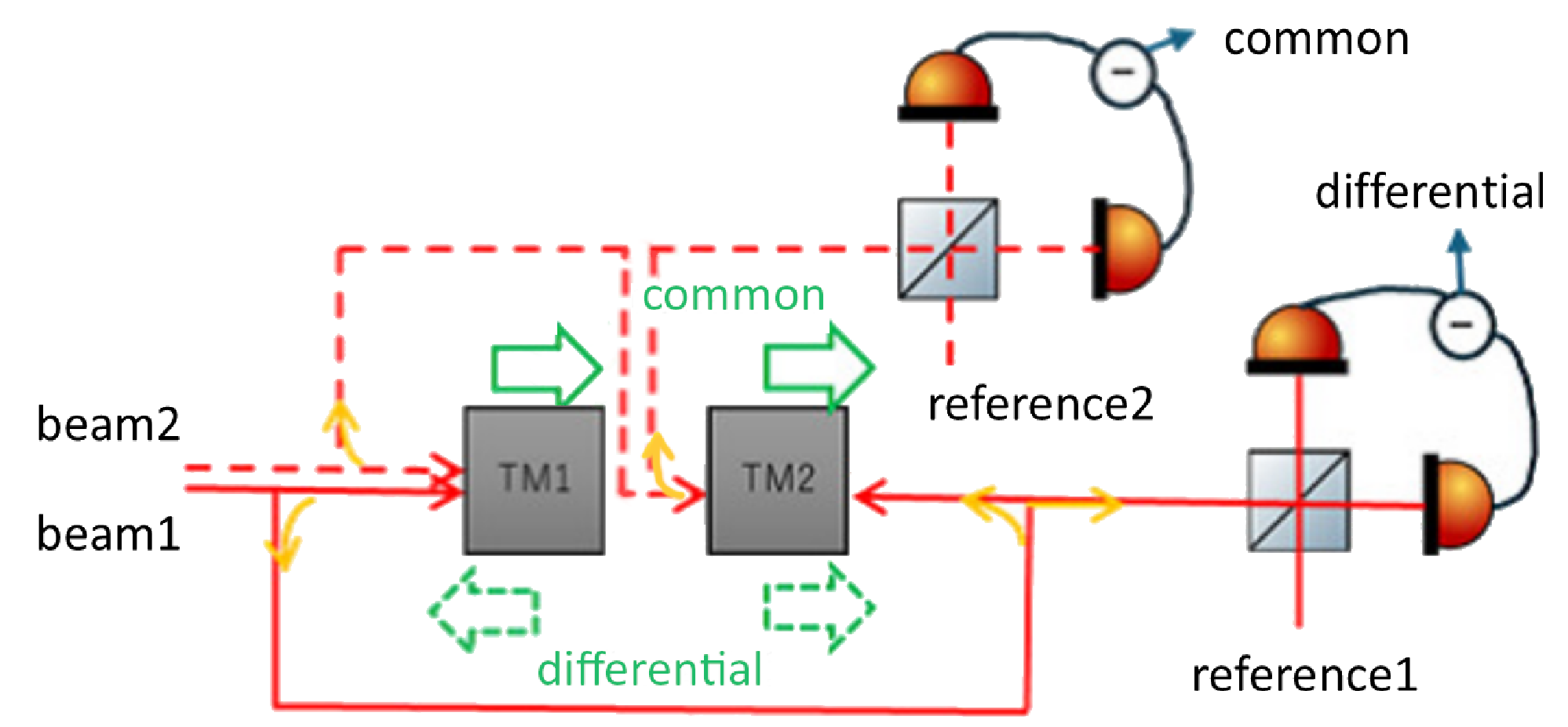}\\
    \caption{ Experimental setup. The differential and common-mode of test mass 1 (TM1) and test mass 2 (TM2) are measured by laser interferometry with a wavelength of $1064\ \mathrm{nm}$.
}
    \label{fig:setup}
\end{figure}

{\it Theory---}\label{sec:setup}
We consider a system where two mirrors (TM1 and TM2) of equal mass $m$ are coupled via gravity. 
Two laser beams are directed at the test masses as shown in Fig.~\ref{fig:setup}. 
The results of the homodyne measurement are 
\begin{align}
    %{X}_{\pm}
    %&=
    %-{x}^{\text{in}}_{\pm}\\
    %{Y}_{\pm}
    %&=-\alpha_\pm{q}_{\pm}
    %{y}^{\text{in}}_{\pm},
    %\bm{z}_\pm=\bm{C}_\pm\bm{x}_\pm+\bm{v}_\pm
     Y_\pm = \bm{C}_\pm\bm{x}_\pm+y_\pm^{\rm in}, 
        \label{linearmeasurement}
\end{align}
where $Y_\pm$ represent the phase quadratures, $\bm{C}_\pm = (-\alpha_\pm, 0, 0)$ denote the optomechanical coupling constants, and $\bm{x}_\pm = (q_\pm, p_\pm, r_\pm)^T$ denote the position, momentum, and auxiliary variables used to describe feedback cooling (for details see Appendix~\ref{app:A}). 
The terms $y^{\rm in}_\pm$ represent vacuum (white) noise, with variances given by $\langle (y_\pm^{\rm in})^2 \rangle = 1$. 
In addition, we introduce vacuum noise $x^{\rm in}_\pm$ for the amplitude quadratures. Throughout this work, the subscripts $+$ and $-$ refer to the common mode and the differential mode, respectively. 
The optomechanical coupling constants $\alpha_\pm$ are given by $\sqrt{16\omega_cP_{\rm in}/(m\Omega_\pm c^2})$,
%\begin{align}
%    %g=\sqrt{\frac{2\omega_cP}{m\Omega c\ell}},\quad
%    \alpha_\pm=\sqrt{\frac{8\omega_cP_{\rm in}}{m\Omega_\pm c^2}},
%    \label{interferometercoupling}
%\end{align} 
where $\omega_c$ is the laser frequency and $P_{\rm in}$ the incident laser power, $m$ the mass of each mirror, $\Omega_+=\Omega$ and $\Omega_-=\Omega\sqrt{1-\delta}$ the resonance frequency of the common and differential modes. The term $\delta=\Lambda\times4Gm/(\Omega^2 L^3)$ represents the gravitational coupling between the two mirrors, where $\Omega$ is the resonance frequency of each mirror, $L$ is the average distance between their centers of mass, and $\Lambda$ is a form factor that accounts for the deviation from a point-mass approximation (see Appendix~\ref{app:B}).
%In this paper, we assume $m/L^3=\rho \Lambda$, where $\rho$ is the mass density of the mirror and $\Lambda$ is the factor determined by the shape and configuration of the mirrors~\cite{Miao}. 

The motion of the mirrors is given by  
\begin{align}
%    \dot{{q}}_\pm&=\Omega_\pm p_\pm,\\
%    \dot{{p}}_\pm&=-(\Omega_\pm+g_0\omega_{\rm fb}) q_\pm-\Gamma p_\pm+\sqrt{2\Gamma} p_\pm^{\rm in}
%    -\alpha_\pm x_\pm^{\rm in}\notag\\
%    &+\frac{g_0 \omega_{\rm fb}}{\alpha_\pm} {y}^{\rm in}_\pm - g_0 \omega_{\rm fb} r_\pm,\\
%    \dot{r}_{\pm}&=-\omega_{\rm fb}r_{\pm}-\omega_{\rm fb}{q}_\pm+\frac{\omega_{\rm fb}}{\alpha_{\pm}}{y}^{\rm in}_{\pm},
\dot{\bm{x}}_\pm=\bm{A}_\pm \bm{x}_\pm+\bm{w}_\pm. 
         \label{eom}
\end{align}
Here, \( \bm{A}_\pm = \begin{bmatrix} 0 & \Omega_\pm & 0 \\ -\Omega_\pm-g_0^\pm\omega_{\rm fb}^\pm & -\Gamma & -g_0^\pm\omega_{\rm fb}^\pm \\ -\omega_{\rm fb}^\pm & 0 & -\omega_{\rm fb}^\pm \end{bmatrix} \) parameterize the state-space model, where $\Gamma$ is the mechanical dissipation rate, $\omega_{\rm fb}^\pm$ are the cutoff frequency of the highpass filter and $g_0^\pm$ are the feedback gain. 
The noise terms are given by $\bm{w}_\pm=\left(0, \sqrt{2\Gamma}p^{\rm in}_\pm-\alpha_\pm x^{\rm in}_\pm+\frac{g_0^\pm \omega_{\rm fb}^\pm}{\alpha_\pm}y^{\rm in}_\pm, \frac{\omega_{\rm fb}^\pm}{\alpha_\pm}y^{\rm in}_\pm\right)^T$, where ${p}^{\rm in}_\pm$ represent thermal (white) noise with variances $\langle({p}^{\rm in}_\pm)^2\rangle = 2n_\pm + 1$, and $n_\pm $ are the average phonon occupation number.    

In a linear quantum measurement described by Eqs.~(\ref{linearmeasurement}) and (\ref{eom}), the signals $Y_\pm$ inherently include shot noise, denoted by ${y}_\pm^{\rm in}$, while the mirror positions $q_\pm$ are intrinsically perturbed by radiation pressure noise, represented by $\alpha_\pm {x}_\pm^{\rm in}$. Based on the results of the measurements, the Kalman filter produces the optimal estimate $\hat{\bm{x}}$, which can be used to analyze the GIE, as follows~\cite{10.1115/1.3658902,Belavkin1980}: 
\begin{align}
\dot{\hat{\bm{x}}}_\pm&=\bm{A}_\pm\hat{\bm{x}}_\pm+\bm{K}_\pm(Y_\pm-\bm{C}_\pm\hat{\bm{x}}_\pm)\label{eq:kalman}\\
\dot{\bm{V}}_\pm&=\bm{A}_\pm \bm{V}_\pm+\bm{V}_\pm \bm{A}_\pm^T+\bm{N}_\pm\notag\\&-(\bm{V}_\pm \bm{C}_\pm^T+\bm{L}_\pm)(\bm{V}_\pm \bm{C}_\pm^T+\bm{L}_\pm)^T, 
\label{eq:riccati}
\end{align}
where $\bm{K}_\pm=\bm{V}_\pm \bm{C}_\pm^T+\bm{L}_\pm$ are the Kalman gain and $\bm{V}_\pm = \langle (\bm{x}_\pm - \bm{\hat{x}}_\pm)(\bm{x}_\pm - \bm{\hat{x}}_\pm)^T \rangle$ are the conditional covariance. In addition, $\bm{N}_\pm=\langle\bm{w}_\pm\bm{w}_\pm^T\rangle$ and $\bm{L}_\pm=\langle y^{\rm in}_\pm\bm{w}_\pm\rangle$. 
Eqs~(\ref{eq:kalman}) and (\ref{eq:riccati}) compute state estimate $\hat{\bm{x}}_\pm$ using a recursive predict-correct cycle. First, the Riccati equation~(\ref{eq:riccati}) determines the theoretical conditional covariance $\bm{V}_\pm$ and the optimal gain $\bm{K}_\pm$, a calculation carried out independently of the measurement data. This predetermined gain is then used to fuse measurements to correct the state prediction, yielding minimum-variance estimate.

%\begin{table}[b]
%    \centering
%    \begin{tabular}{c|c}
%    \hline\hline
%~~~~$\Gamma/2\pi$~~~~ & $10^{-18}~{\rm Hz}$
%\\ \hline
%~~~~~~$\Omega/2\pi$~~~~~~ & ~~~~~~$10^{-3}~{\rm Hz}$~~~~~~
%\\ \hline
%$\gamma_m/2\pi$& $10^{-6}~{\rm Hz}$
%\\ \hline
%$\omega_c/2\pi$ & $2.8\times10^{14}~{\rm Hz}$
%\\ \hline
%${P}$& $10^{-2}~{\rm mW}$
%\\ \hline
%$\ell$ & $10~{\rm cm}$
%\\ \hline
%$T$ & $2.7~{\rm K}$
%\\ \hline
%$\rho$ & $ ~~~~20~{\rm g}/{\rm cm}^3$~~~~
%\\ \hline
%$m$ & $100~{\rm mg}$
%\\ \hline
%$\Lambda$ & $2$
%\\ \hline
%${\Delta/\kappa}$ & $0.05$
%\\ 
%\hline
%    \end{tabular}
%    \caption{
%    Parameters for the demonstration in GIE generation.
%    }
%    \label{tab:my_label}
%\end{table}

%%%%%%%%%%%%%%%%%%%%%%%%%%%
%\section{Gravity-induced entanglement generation}
%%%%%%%%%%%%%%%%%%%%%%%%%%%

The Gaussian state entanglement can be characterized by the entanglement negativity defined by

\begin{eqnarray} 
E_{N}=
    -\frac{1}{2}\log_2\left[\frac{\Sigma-\sqrt{\Sigma^{2}-4\text{det}\bm{\mathcal{V}}}}{2}\right]. 
    \label{negativity}
    \end{eqnarray}
Here, $\bm{\mathcal{V}}$ is the conditional covariance matrix of the individual test masses (see Appendix~\ref{app:C}) and $\Sigma=\det\bm{\mathcal{V}}_{1}+\det\bm{\mathcal{V}}_{2}-2\det\bm{\mathcal{V}}_{12}$, where $\bm{\mathcal{V}}_1$, $\bm{\mathcal{V}}_2$
are the covariance matrices of TM1 and TM2 normalized by the frequency $\Omega$, respectively. $\bm{\mathcal{V}}_{12}$ represents the gravity-induced correlation matrix between individual masses. 
According to the separability condition for two-mode Gaussian states, the systems are entangled if and only if $E_N>0$~\cite{Giedke2001}.
To verify whether quantum entanglement is present from Eq.~(\ref{negativity}), it is sufficient to confirm $\Sigma-\text{det}\bm{\mathcal{V}}>1$ for $\Sigma>2$. 

\begin{figure}
    \centering
    \includegraphics[width=8.cm]{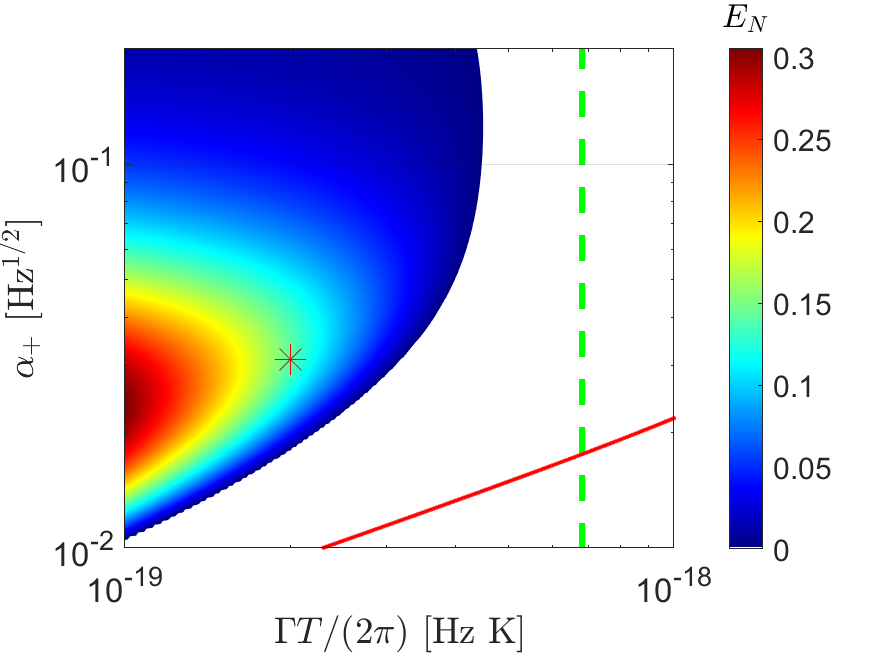}\\
%    \includegraphics[width=8.25cm]{SNR=1ver2.pdf}\\
    %\vspace{5mm}
    %\includegraphics[width=8.cm]{timeyalphagt.pdf}
    \caption{Contour plot of $E_N$ in the $\alpha_+$-$\Gamma T/(2\pi)$ plane. The green dashed line indicates the threshold for generating the GIE in Eq.~\eqref{CoAB}, the red curve shows the boundary where the relaxation time due to the Kalman filter is equal to half the thermal decoherence time (see Appendix~\ref{app:A}). The star symbol corresponds to the parameters used in the simulation. The GIE can be generated when $\Gamma T/(2\pi)<4.5\times10^{-19}$\ Hz\ K.
    %Contour of total experimental time to achive S/N=1 on the plane of $P$ - $T\Gamma/2\pi$ (lower panel). 
%These panels assume 
%the parameters listed in Table I. 
}
    \label{fig:enter-label}
\end{figure}

The colored region in Fig.~\ref{fig:enter-label} shows the region where GIE is generated.  
The generation rate derived in Ref.~\cite{Krisnanda2020} must be faster than the thermal decoherence rate, as shown below:
    \begin{eqnarray}
    &&2\Gamma (2n_++1)<\delta\Omega. 
    \label{CoAB}
%    \\
%&&2\Gamma\left(2n_{\rm th}^++1\right)<    .
%    \label{CoB}
\end{eqnarray}
From the above inequality, the condition for the GIE is bounded by 
%\begin{eqnarray}
%{k_{\rm B}T\Gamma\over \hbar G(m/L^3)}< 1,
%\label{encrr}
%%\end{eqnarray}
%which is rephrased as
\begin{eqnarray}
    %{\left({T\over  2.7{\rm K}}\right)}
        {\left({
    \Gamma T/2\pi\over 6.5\times10^{-19}{\ \rm Hz\cdot K}}\right)}
     {\left({20000{\ \rm kg/m^3} \over \rho }\right)}
     \left({1\over \Lambda}\right)<1,
%     \left(\frac{1}{\sqrt{1+\zeta^2+\bar{n}_+}+\zeta}\right)<1,
         \label{ConditionE}
\end{eqnarray} 
where $T$ represents the temperature of the test masses and $\rho$ is the mass density of the mirror. 
%The required $\Gamma T/(2\pi)$ is about $10^{-18}\,\mathrm{Hz\cdot K}$. 
 %, and in the next section, we will discuss whether this value is achievable.

%%%%%%%%%%%%%%%%%%%%%%

{\it Sources of decoherence---}\label{sec:feasibility}
As inherently unavoidable sources of decoherence, we consider gas damping, black-body radiation, and cosmic-ray collisions. 
First, the dissipation rate caused by gas, involving particles with mass $m_{\rm atom}$, pressure $P$, and temperature $T_{\rm env}$ is given by~\cite{2010PhLA..374.3365C} 
\begin{eqnarray}
{\Gamma_{\rm gas}}
&=&{PR^2\sqrt{8\pi m_{\rm atom}}\over m\sqrt{k_BT_{\rm env}}}{ }\left(1+{h\over 2R}+{\pi\over4}\right),
%\nonumber\\
%&=&3\times 10^{-15}\times 2\pi\left(
%{P\over 3\times 10^{-13}{\rm Pa}}\N[right)
%\left(
%{m\over  0.01{\rm kg}}\right)^{-1}
%\left(
%{m_{\rm nitrogen~molecule}\over 4.7 \times 10^{-26}{\rm kg}}\right)^{1/2}
%\left(
%{T\over 10^{-3}{\rm K}}\right)^{-1/2}
%{\rm Hz}
%\nonumber
%\\
%&=&0.96\times 10^{-18}{\rm Hz}\left(
%{P\over 1\times 10^{-14}{\rm Pa}}\right)
%\left(
%{m\over  100{\rm mg}}\right)^{-1/3}
%\nonumber\\
%&&\hspace{-8mm}\times
%\left({20 {\rm g/cm^3}\over \rho}\right)^{-2/3}
%\left(
%{m_{\rm atom}\over 1.7 \times 10^{-27}{\rm kg}}\right)^{1/2}
%\left(
%{T\over 2.7 {\rm K}}\right)^{-1/2}. 
%\nonumber\\
%&\simeq&10^{-18}{\rm Hz}
%{P\over 1.2\times10^{-13}\ {\rm Pa}}
%\left(
%{1\ {\rm kg}\over m }\right)^{\frac{1}{3}}
%\left(
%{1.7\ {\rm K}\over T}\right)^{\frac{1}{2}}
\label{residualgas}
%&=&0.88\times 10^{-16}\times 2\pi\left(
%{P\over 1\times 10^{-13}{\rm Pa}}\right)
%\left(
%{m\over  100{\rm g}}\right)^{-1/3}
%\left({20 {\rm g/cm^3}\times (\pi/4)\over \rho\Lambda}\right)^{2/3}
%\nonumber\\
%&&\times
%\left(
%{m_{\rm atom}\over 1.7 \times 10^{-27}{\rm kg}}\right)^{1/2}
%\left(
%{T\over {\rm mK}}\right)^{-1/2}
%{\rm Hz}
\end{eqnarray} 
where $k_B$ is the Boltzmann constant, and $h$ and $R$ denote the height and radius of the mirror, respectively. 

Second, following~\cite{Rijavec:2020qxd}, the total decay rate from thermal blackbody photons is estimated as $ \Gamma_{\rm scat} + \Gamma_{\rm abs} + \Gamma_{\rm em}\,$where we define 
%\begin{eqnarray}
%{\tilde\Gamma_{\rm scat}+\tilde\Gamma_{\rm abs}+\tilde\Gamma_{\rm em}\over 4a^2}{\hbar^2\over mk_BT},
%end{eqnarray}
%where $\tilde\Gamma_{\rm scat}$, $\tilde\Gamma_{\rm abs}$, $\tilde\Gamma_{\rm em}$ denote the decoherence rate for the scattering, absorption, and emission processes, respectively, and $a=\pi^{2/3}\hbar c/(2k_BT)$, where used 
\begin{align}
%  \raisetag{-0.5\normalbaselineskip}
  \Gamma_{\rm scat}
  &={8!{8}\hbar R^6\over  9\pi m}\left({k_BT_{\rm env}\over\hbar c}\right)^8\zeta(9){\rm Re}\left[{\varepsilon -1\over \varepsilon+2}\right]^2
   \label{radiationscat}
\end{align}
for the scattering by thermal photons, and 
\begin{align}
 % \raisetag{-0.5\normalbaselineskip}
  \Gamma_{\rm abs} 
  &= 
     {16\pi^{5}
     \hbar R^3\over 189m} \,  
     \left({k_BT_{\rm env}\over\hbar c}\right)^5 
     {\rm Im}\!\Biggl[\frac{\varepsilon - 1}{\varepsilon + 2}\Biggr]
  \label{radiationemit}
\end{align}
for absorption. Here, $\hbar$ is the reduced Planck constant, $\varepsilon$ is the dielectric constant, and $\zeta(z)$ is the Zeta function. 
For emission, $\Gamma_{\rm em}$ is obtained by replacing $T_{\rm env}$ in Eq.~(\ref{radiationemit}) with the temperature of the test masses.

Finally, we simulate the acceleration noise from cosmic-ray collisions using Geant4~\cite{G4} and a flux model~\cite{cosmic-ray-flux}\cite{LPF-flux}. After removing large events through data processing, the requirement is met with a duty cycle—the fraction of usable data time—of 12\% (see Appendix~\ref{app:D}).

{\it Simulation---}
The parameter~$\Sigma - \det(\bm{\mathcal{V}})$ depends on the fourth power of the variance components and therefore does not follow a Gaussian distribution. Consequently, instead of using error propagation, we estimate the integration time required to demonstrate GIE through simulations.
First, the sample paths of the differential and common mode motions (hereafter referred to as the 'true values'), subject to thermal noise, radiation pressure noise, and feedback noise, are computed using the Euler--Maruyama method~\cite{Maruyama1955ContinuousMP} with a sampling rate of 10\,Hz.  
Second, the measurement data \(Y_\pm\) are generated by multiplying these paths by the optomechanical coupling constants \(\alpha_\pm\) and adding shot noise. 
Finally, the optimal state estimates \(\hat{\bm{x}}\) are obtained by applying the Kalman filter, defined in Eqs.~(\ref{eq:kalman}) and (\ref{eq:riccati}), to the simulated measurement data \(Y_\pm\).

In the simulation, we set $\Omega_+/(2\pi) = 2.5\times10^{-4}\ \mathrm{Hz}$, $\alpha_+=0.031\ \mathrm{Hz^{1/2}}$, $\Gamma T/(2\pi) = 2\times10^{-19}\ \mathrm{Hz\cdot K}$, $\rho = 20\ \mathrm{g/cm^3}$, $g_0^+ = 0.2$, $\omega_{\rm fb}^+ = 5\Omega_+$, $g_0^- = 1$, and $\omega_{\rm fb}^- = 2\Omega_-$. 
%, corresponding to the star symbol in Fig.~\ref{fig:enter-label}.
For a cylindrical test mass with thickness $h$, radius $R = 0.5 h$, and center-of-mass separation $L = 1.25h$, we find that $\delta = 0.91$ and $\Sigma-\det(\bm{\mathcal{V}}) = 1.27$. For a 1 kg test mass, the zero-point amplitude is $q^+_{\rm zpf}=1.8\times10^{-16}$~m and the surface-to-surface gap is $h/4\simeq1$~cm. 
The required resonant frequency of $\Omega_+/(2\pi) = 2.5\times10^{-4}\ \mathrm{Hz}$ can be achieved either passively, through the gravitational field of ancillary tuning masses, or actively, using an electrostatic feedback system. 

Since the above parameters are configured to keep the feedback noise sufficiently low, we can extract the shot noise level, the combined thermal and radiation pressure noise level, and the effective susceptibility by fitting the measurement data to the following expression: 
\begin{eqnarray}
S_{\rm fit}^{\pm}(\omega) = S^{\pm}_{\rm tot} |\chi^{\pm}_{\rm eff}(\omega)|^2 + S^{\pm}_{\rm bg}, 
\end{eqnarray}
where $S^{\pm}_{\rm bg}$ are the spectral background floor, $S^{\pm}_{\rm tot}$ the sum of force noise, $\chi^{\pm}_{\rm eff} = \Omega_\pm/(\omega^2-\Omega_{\pm}^2+i\omega\Gamma+g_0^\pm\omega_{\rm fb}^\pm\Omega_{\pm}i\omega/(i\omega+\omega_{\rm fb}^\pm))$ the effective mechanical susceptibility modified by the feedback. 
%The weighted non-linear least-squares fitting is performed using the lsqnonlin function in MATLAB's optimization toolbox. To improve accuracy, the fitting procedure is repeated 100 times using the MultiStart option in the Global Optimization Toolbox.
Using these fitted parameters, we numerically solve the Lyapunov equation - given by the first line of Eq.~(\ref{eq:riccati}) - to compute the unconditional variances, $\bm{V}_\pm^{\rm un} = \langle \bm{x}_\pm \bm{x}_\pm^T \rangle$. %, while setting the off-diagonal components to zero. 
The difference between the unconditional variances and the estimated variances $\hat{\bm{V}}_\pm=\langle\hat{\bm{x}}_\pm \hat{\bm{x}}_\pm^T\rangle$ yields the conditional variances $\bm{V}_\pm$, according to Eve's law~\cite{BlitzsteinHwang2014}. 

\begin{figure}
    \centering
 \includegraphics[width=9cm]{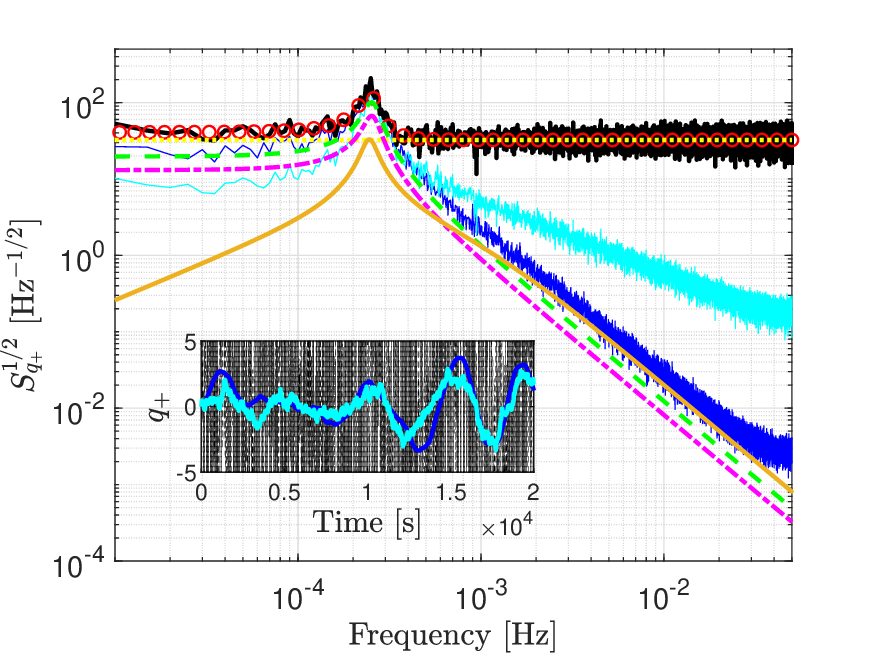}\\
    %\vspace{5mm}
    %\includegraphics[width=8.cm]{timeyalphagt.pdf}
    \caption{An example of the common-mode ASD \(S_{q_{+}}^{1/2}\), normalized by the zero-point amplitude $q^+_{\rm zpf}$, measured over \(3\times10^{6}\) seconds. For a 1 kg test mass, the value 1 on the vertical axis corresponds to $1.8\times10^{-16}\ {\rm m/\sqrt{Hz}}$. 
Blue shows the true values, light blue the estimate, black the measured data.  
Yellow is shot noise, green is radiation pressure noise, magenta is thermal noise, orange is feedback noise, and red circles show fit results.  
The inset shows the time-series data \( q_+ \) from 0 to $2\times10^4$ seconds, with colors matching the main figure.  
Due to shot noise, the measured signal has an amplitude of about 20, causing the inset’s background to appear black.}
    \label{fig:simulation}
\end{figure}

An example of the common mode amplitude spectral density (ASD) is shown in Fig.~\ref{fig:simulation}. 
The ASD is obtained using Welch’s method~\cite{1161901}, with 50\% overlap and a Hanning window. 
From about 100 simulation runs, we compute the probability density functions of \( \Sigma - \det(\bm{\mathcal{V}}) \), \( \bm{V}_+ \), and \( \bm{V}_- \), with four integration times: \( 10^5 \)~s, \( 5\times10^5 \)~s, \( 2\times10^6 \)~s , and \( 3\times10^6 \)~s (for details see Appendix~\ref{app:F}).

{\it Results and Discussion---} 
Let the theoretical value of
\(\Sigma - \det(\bm{\mathcal{V}})\) be denoted by \(\beta\), and define \(\Delta = \beta-1.\)
Then the probability of observing GIE can be given by:
\begin{align}
      P_{\mathrm{obs}}^{\mathrm{GIE}}
  = \int_{1}^{1+2\Delta} p\bigl(\ \Sigma - \det(\bm{\mathcal{V}})\bigr)\,\mathrm{d}\bigl(\Sigma - \det(\bm{\mathcal{V}})\bigr)
%  = \Pr\bigl[\,1-\Delta \le \Sigma - \det(V) \le 1\,\bigr],
\end{align}
where \(p(\Sigma - \det(\bm{\mathcal{V}}))\) is the probability density function.
%In the lower panel of Fig.~\ref{fig:histogram}, star symbols denote \(P_{\mathrm{obs}}^{\mathrm{GIE}}\), including the case with measurement time \(1\times10^5\)\ s.
For each measurement time, the cumulative probabilities \( P_{\text{obs}}^{\text{GIE}} \) are 13\%, 34\%, 79\%, and 88\%, respectively. 
As the result demonstrates an improvement proportional to the square root of the measurement time, extrapolating it suggests that the measurement time required to achieve a probability of 99.7\% is \( 3.5\times10^6 \)~s. 
Although the measurement time becomes one year when the duty cycle of 10\% is taken into account for 1 kg test masses, it is the same as the operational period of LPF. 

For comparison, we calculate the values of $P_{\text{obs}}^{\text{GIE}}$, for $\delta = 0.75$, which corresponds to the distance of $L=4h/3$. 
%in Fig.~\ref{fig:histogram} as circle symbols. 
In this case, the value of $\Sigma-\det(\bm{\mathcal{V}})$ is 1.09, and compared to the value of 1.27 at $\delta = 0.91$, there is a 3-fold difference in $\Delta$. Therefore, the integration time required to achieve 99.7\% confidence is expected to differ by a factor of $\sim10$ (see Appendix~\ref{app:F}). 
The result supports this prediction.

\begin{figure}
    \centering
 \includegraphics[width=8.cm]{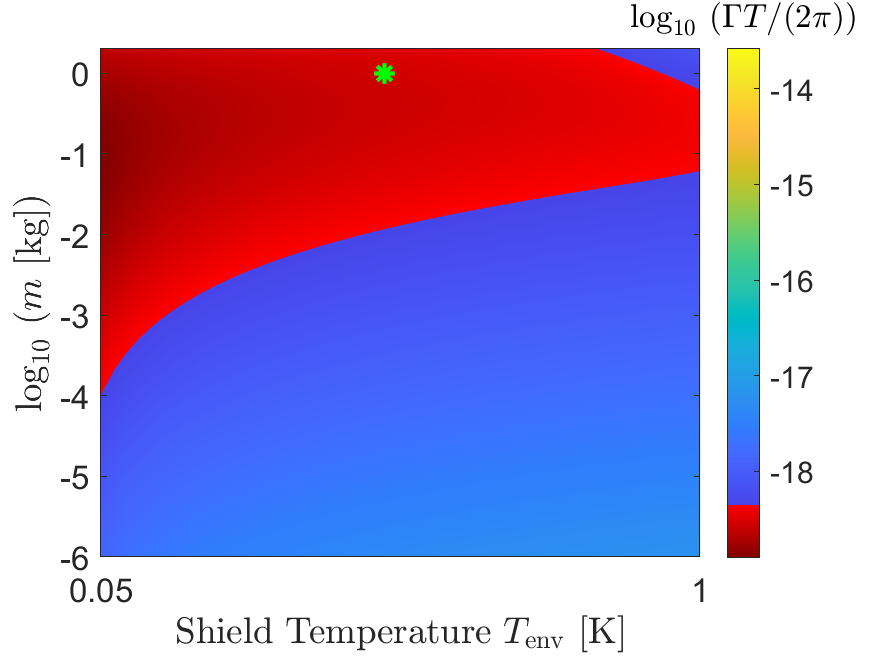}\\
    %\vspace{5mm}
    %\includegraphics[width=8.cm]{timeyalphagt.pdf}
    \caption{Contour plot of $\log(\Gamma T/(2\pi))$ in the $\log m$-$T_{\rm env}$ plane at $\alpha_+=0.031\ \mathrm{Hz^{1/2}}$. %The red region indicates the area that meets the requirements for the GIE.
    }
    \label{fig:gammaT}
\end{figure}

In terms of decoherence, the red region in Fig.~\ref{fig:gammaT} shows the region where the total dissipation given by Eqs.~(\ref{residualgas}), (\ref{radiationscat}), and (\ref{radiationemit}) satisfies
$\Gamma T/(2\pi) \leq 4.5\times10^{-19}\ \mathrm{Hz\cdot K}$. Here, we consider a cylindrical mass with 3\% absorption at a wavelength of 1064 nm~\cite{Shiraishi2001,PhysRevB.6.4370}, with radius $R = 0.5 h$ and center-of-mass separation $L = 1.25h$. 
These test masses are surrounded by a shield at temperature $T_{\rm env}$. 
We assume that the residual gas is an ideal gas consisting of hydrogen at a pressure of $5\times10^{-15}\ \mathrm{Pa}$ at 4 K. The test masses, with an emissivity of 0.04 \cite{PhysRev.28.174}, are cooled by radiation. 

By increasing the mass of the test masses, the gas damping can be mitigated. 
However, this comes at the cost of increased decoherence because of blackbody radiation emission. This trade-off is important in vacuum, where the levitated mass can only dissipate heat via radiation. As the mass increases, the incident laser power must be increased to maintain the value of $\alpha_\pm$, causing radiative cooling to become less effective. Consequently, the equilibrium temperature increases. For large masses, blackbody radiation becomes more influential, whereas for small masses, gas damping becomes more significant. As a result, the optimal mass scale for the GIE is on the order of grams to kilograms. For example, when a 1~kg test mass is placed inside a shield cooled to $T_{\rm env}=0.1$~K and cooled by radiation, it attains an equilibrium temperature of $T=11.7$~K under an absorbed laser heating power of $2\times0.14$ uW. This results in a value of $\Gamma T/(2\pi) = 2.0 \times 10^{-19}\ \mathrm{Hz\cdot K}$.

In principle, three key conditions must be validated before realizing the proposed space‐based experiment.  
%First, to observe the GIE, the gravitational interaction between the test masses must dominate over all other forces. Since gravitational coupling between milligram‐scale objects has already been demonstrated~\cite{Westphal2021}, this condition can be satisfied. 
%Second, it is essential to demonstrate the generation of quantum entanglement between macroscopic objects through continuous measurement. While conditional entanglement has been realized in spin systems~\cite{SpinRef}, it remains crucial to experimentally verify such entanglement between massive test masses on the ground, independent of the interaction mechanism.  
%Ground‐based experiments are inherently limited by decoherence due to seismic noise and gas damping. Nevertheless, reproducing GIE‐like conditions via stronger interactions such as radiation pressure~\cite{Miki22} offers a viable approach to experimentally confirming the underlying principles of quantum entanglement generation.  
First, achieving extremely high vacuum ($5\times10^{-15}$ Pa) and cryogenic temperatures ($\sim$0.1 K) in the space environment requires precise engineering design and quantitative validation. Although the required vacuum level is challenging, the ground-based experiment has already achieved $5\times10^{-15}$ Pa at 4.2 K by cryopumping~\cite{PhysRevLett.89.233401}. %For the space mission, a 1 K–class Joule–Thomson cryocooler~\cite{SATO201647}—with a lifetime exceeding 3 years and a nominal cooling power of 10 mW at 1.7 K—is suitable. A cryocooler capable of cooling down to 0.05\ K has also been developed~\cite{PROUVE2020103144}. 
For space missions, a cryocooler capable of cooling down to 0.05 K has been developed~\cite{PROUVE2020103144}. This Adiabatic Demagnetization Refrigerator (ADR) provides a nominal cooling power of 0.4 µW at 0.05 K, which is greater than the heat input from the laser. 
A spherical cryogenic shield (e.g., ~20 cm inner diameter, ~50 kg) would by itself create a resonant frequency of 0.29 mHz. To counteract this and passively set the frequency to within a few percent of the target, two external tuning masses (e.g., 88 kg each at 30 cm) provide a counteracting gravitational gradient. Any residual offsets can then be precisely corrected using electrostatic feedback. Importantly, a custom launch-lock mechanism is required to operate in the cryogenic, ultra-high vacuum environment. 

Second, unlike LPF, this study must precisely measure not only the differential motion of the test masses but also their common-mode motion. In LPF, the spacecraft is controlled with respect to one test mass (drag-free control) to mitigate disturbances such as the solar wind; however, the achievable stability is limited by noise from the thrusters used for control, which leads to elevated noise levels in the common-mode motion. To avoid this, noise reduction via passive shielding is effective (for details see Appendix~\ref{app:G}). 

Finally, acceleration noise from patch effects remains a significant challenge that requires further investigation, particularly at cryogenic temperatures and low frequencies. This noise arises from electrostatic forces between adjacent metallic surfaces—such as those between test masses and actuators—and is driven by fluctuations in both charge and surface patch potentials. Its behavior is frequency-dependent: high-frequency components have been observed to decrease significantly under cryogenic conditions~\cite{PhysRevLett.101.180602}, whereas its low-frequency characteristics are not yet well characterized. 

Effective mitigation strategies include general approaches such as shielding and charge neutralization. Moreover, the noise level can be further reduced through careful design, as the patch force itself decreases with greater surface separation, while the resulting acceleration is inversely proportional to the test mass. Our design takes advantage of both principles by incorporating a large (1 kg) test mass and a wide (1 cm) separation between surfaces, and is therefore expected to be highly effective in suppressing this noise source.

{\it Summary---}\label{sec:conclusions}
Experimental investigation of the quantum nature of gravity is crucial for advancing modern physics, yet it remains a significant challenge. In this study, we show that such an investigation is marginally achievable with current technology. To this end, we simulated quantum measurements of two adjacent kg-scale test masses separated by a cm-scale distance. 
Our simulation results indicate that a detectable signature—Gravitationally Induced Entanglement (GIE)—can be generated between the masses. Achieving an experimental demonstration of this GIE, however, requires stringent conditions. These include attaining the highest vacuum levels possible in terrestrial experiments, cooling the test masses inside state-of-the-art refrigerator. Although a deeper understanding of low-temperature, low-frequency patch potentials is necessary to demonstrate the GIE, mitigating this noise source appears feasible, suggesting that experimental observation of GIE is within reach.

{\it Acknowledgement---} 
We thank Markus Arndt and Ichiro Arakawa for valuable discussions, and Koichi Ichimura and Tomoki Fujii for providing the Geant4 source codes.
N.M. conceived and designed the research, performed simulations with K.S., and wrote the manuscript.
K.S. developed the state-space model, and K.H. carried out the GEANT4 simulations. %D.M. and K.Y. derived the GIE condition (Eq. 6). --> add citation of PhysRevD.110.024057 at Eq6. 
All authors discussed the results.
N.M. is supported by JST FOREST Grant No.~JPMJFR202X. 
K.S. is supported by RIKEN TRIP initiative (RIKEN Quantum).
K.Y. and A.M. are supported by JSPS KAKENHI Grant No.~JP23H01175, and D.M. is supported by JSPS Overseas Research Fellowships.

\appendix

\setcounter{equation}{0}
\renewcommand{\theequation}{A\arabic{equation}}
\section{State space model}
\label{app:A}
Here, we present the detailed derivation of Eqs. \eqref{linearmeasurement} and \eqref{eom}. 
These equations can also be applied to a dark-fringe Michelson interferometer with test masses in both arms. 

First, let us denote the complex amplitude of the laser incident at \(q_1=0\) by \(\bar a + a^\text{in}\),  
where \(\bar a\) is the mean value and \(a^\text{in}\) is the vacuum fluctuation.  
We normalize it so that $|\bar a|^2$ equals the mean photon flux $P_\text{in}/(\hbar\omega_c)$, and without loss of generality, we take \(\bar a\) to be real: $\bar a = \sqrt{P_\text{in}/(\hbar\omega_c)}$. 
The laser amplitude reflected from TM1 at a general position \(q_1\) is, when referenced to \(q_1=0\),
\begin{align}
  a &= -\,\bar a\,e^{2ikq_\text{zpf}q_1} \;+\; a^\text{in},
  \nonumber\\
  &\simeq -\,\bar a \;-\;2ikq_\text{zpf}\,\bar a\,q_1 \;+\; a^\text{in}, 
\end{align}
where $k=\omega_c/c$ and $q_\text{zpf}=\sqrt{\hbar/(2m\Omega)}$. 
The phase quadrature of the light reflected from TM1, given by \(Y_1 = -\,i\,(a - a^*)\), is
\begin{align}
  Y_1 &= -\,4kq_\text{zpf}\,\bar a \,q_1 \;+\; y^\text{in}
      \nonumber\\
    &= -\,(\alpha/\sqrt{2}\,)\,q_1 + y^\text{in},
\end{align}
where $\alpha = \sqrt{16\omega_c P_\text{in}/(m\Omega c^2)}$. Then, the laser is reflected off TM2, either on the same side or on the opposite side, depending on whether the laser is used to detect the common mode or the differential mode, respectively.
After reflection, the phase quadrature is
\begin{align}
Y_\pm = \mp\,(\alpha/\sqrt{2}\,)\,q_2+ Y_1,
\end{align}
where the upper and lower sign correspond to the common and differential modes, respectively.
Thus, we recover the observation equation Eq. \eqref{linearmeasurement}, with the following definition of the mechanical variables of the two modes:
\begin{align}
q_\pm=\sqrt{\frac{\Omega_\pm}{\Omega}}\frac{q_1\pm q_2}{\sqrt{2}},
~~~~
p_\pm=\sqrt{\frac{\Omega}{\Omega_\pm}}\frac{p_1\pm p_2}{\sqrt{2}}.
\label{modes_definition}
\end{align}

Second, to derive the equation of motion, Eq.~\eqref{eom}, we first consider the case without the feedback cooling. In this case, the equation of motion for the test masses are given by
\begin{align}
\dot{q}_i=&\Omega\,p_i~~(i=1,2),\\
\dot{p}_1=&-\Omega\,q_1-\Gamma\, p_1+\frac{\delta}{2}(q_1-q_2)\nonumber\\
&+\sqrt{2\Gamma}\,p^\text{in}_1-\frac{\alpha}{\sqrt{2}}(x^\text{in}_++x^\text{in}_-),\\
\dot{p}_2=&-\Omega\,q_2-\Gamma\, p_2-\frac{\delta}{2}(q_1-q_2)\nonumber\\
&+\sqrt{2\Gamma}\,p^\text{in}_2-\frac{\alpha}{\sqrt{2}}(x^\text{in}_+-x^\text{in}_-).
\end{align}
Here, $\alpha x^\text{in}_\pm$ represent the radiation pressure noise. Although they acquire time delay during the propagation between the test masses, it is far smaller than the time scale of oscillator and negligible. By combining the above equations along with Eq.~\eqref{modes_definition}, we obtain
\begin{align}
\dot{q}_\pm=&\Omega_\pm\,p_\pm,\\
\dot{p}_\pm=&-\Omega_\pm\,q_\pm-\Gamma\,p_\pm+\sqrt{2\Gamma}\,p^\text{in}_\pm-\alpha_\pm\,x^\text{in}_\pm.
\label{eom_without_cooling}
\end{align}
Next, we take feedback cooling into account, which motivates us to introduce the auxiliary variable \( r(t) \) as follows. For notational simplicity, we omit the subscripts denoting the modes \(\pm\) here.

The dynamics under feedback control is described by the above equations of motion with an additional force term \( F_{\mathrm{fb}} \) on the right-hand side of Eq.~\eqref{eom_without_cooling}. For feedback cooling, this term is chosen as
\begin{align}
F_\mathrm{fb} = -\,g_0\,\Omega\,\hat{p},
\label{cooling_force}
\end{align}
where $g_0$ is the dimensionless feedback gain, so that cooling adds an effective damping $g_0\,\Omega$.  
We estimate $\hat p(t)$ from the phase-quadrature record $Y(t)$ via a first-order high-pass filter:
\begin{align}
\hat{p}(t)
&= \omega_\mathrm{fb}\int^t \!dt'\,e^{-\omega_\mathrm{fb}(t-t')}
      \frac{\dot{Y}(t')}{-\alpha\Omega}
\label{p_est_cooling}\\
&= -\,\frac{\omega_\mathrm{fb}}{\alpha\Omega}\,Y(t)
  + \frac{\omega_\mathrm{fb}^2}{\alpha\Omega}
    \int^t \!dt'\,e^{-\omega_\mathrm{fb}(t-t')}Y(t'),
\nonumber
\end{align}
using \(\dot{Y}/(-\alpha\Omega)\approx p\) and where \(\omega_\mathrm{fb}\) sets the estimator’s bandwidth.  
Adding Eq.~\eqref{p_est_cooling} with Eq.~\eqref{cooling_force} to \eqref{eom_without_cooling} yields a non-Markov system with colored noise.  
To restore Markovianity, we define
\begin{align}
r(t)
= \frac{\omega_\mathrm{fb}}{\alpha}
  \int^t\!dt'\,e^{-\omega_\mathrm{fb}(t-t')}\,Y(t'),
\label{r_introduced}
\end{align}
whose evolution is
\begin{align}
\dot r
&= -\,\omega_\mathrm{fb}\,r + \frac{\omega_\mathrm{fb}}{\alpha}\,Y
\nonumber\\
&= -\,\omega_\mathrm{fb}\,r - \omega_\mathrm{fb}\,q + \frac{\omega_\mathrm{fb}}{\alpha}\,y^\mathrm{in}.
\end{align}
Together, Eqs.~\eqref{eom_without_cooling}, \eqref{cooling_force}, \eqref{r_introduced}, and this $\dot r$–equation form the Markovian system given in Eq.~\eqref{eom}.  
In the \((q,p,r)\) formulation the feedback noise is treated exactly.  If one attempted cooling by increasing $\Gamma$ in a model with \((q,p)\), the feedback noise would be omitted, which would violate the uncertainty relation for the conditional covariance.
For a general linear system whose operators are governed by the equation of motion [Eq.~\eqref{eom}] and the observation equation [Eq.~\eqref{linearmeasurement}],  
the expectation values and covariances of the conditional state are described 
%by equations that are directly analogous to those in the Kalman filter, 
by the quantum Kalman filter given by Eqs.~\eqref{eq:kalman} and \eqref{eq:riccati}.

Finally, to gain further insight, we present an analytical solution in an effective theory that neglects feedback noise. In the steady state, each element of the conditional covariance matrix is given by the following expressions~\cite{Miki22}: 
\begin{eqnarray}
&&V^\pm_{q q}=\displaystyle{\gamma_\pm-\Gamma\over \alpha_\pm^2}
\\
&&V^\pm_{qp}=\displaystyle{(\gamma_\pm-\Gamma)^2\over 2\alpha_\pm^2\Omega_\pm}
\\
&&V^\pm_{p p}=
\displaystyle{(\gamma_\pm-\Gamma)(2\Omega_\pm^2+\gamma_\pm^2-\Gamma\gamma_\pm)\over 2\alpha_\pm^2\Omega_\pm^2}.
\end{eqnarray}
Here, we introduce the characteristic frequency scales $\gamma_\pm$ of the Kalman filter, which are given by
\begin{eqnarray}
\gamma_\pm=\sqrt{\Gamma^2-2\Omega_\pm^2+2\Omega_\pm\sqrt{\Omega_\pm^2+(2\Gamma(2n_\pm+1)+\alpha_\pm^2)\alpha_\pm^2}}.
\nonumber\\
\end{eqnarray}
%where $\bar{n}_\pm=2\Gamma(2n_\pm+1)+\alpha_\pm^2$. 
Considering the susceptibility of the position including the filter, it is expressed, just as in the original $\Gamma$, by
\begin{eqnarray}
\chi(\omega) &\propto& \frac{1}{\Omega_{\pm}^{2} - \omega^{2} - i\,\omega\,\gamma_{\pm}\ }\,.
\end{eqnarray}
Thus, $\gamma_\pm$ represent the full width at half-maximum (FWHM).
To demonstrate the GIE, the condition $\gamma_+>2\Gamma n_+$ represents an approximate threshold, as depicted by the red curve in Fig.~\ref{fig:enter-label} of the main text. 

\section{Form factor}
\label{app:B}
\renewcommand{\theequation}{B\arabic{equation}}

For two point masses $m$ separated by a distance $L$, the interaction Hamiltonian expanded to the second order in the differential displacement $q$ is:
\begin{align}
    H_{\text{int}} &= \frac{m\Omega^2}{2}q^2 - \frac{Gm^2}{L+q}\notag\\ &\approx \frac{m\Omega^2}{2}q^2 - \left(\frac{Gm^2}{L} - \frac{Gm^2}{L^2}q + \frac{Gm^2}{L^3}q^2\right)
\end{align}
The term proportional to $q^2$ modifies the effective spring constant of the system. We can express this modification using the dimensionless gravitational coupling parameter $\delta$:
\begin{equation}
    H_{\text{int}} \approx \frac{m\Omega^2(1-\delta)}{2}q^2, \quad \text{where} \quad \delta = \frac{4Gm}{\Omega^2 L^3}
\end{equation}
for the point-mass case.

However, our experimental setup uses mirrors of finite cylindrical size. To account for the deviation from the point-mass approximation, we introduce a form factor $\Lambda$. This factor modifies the gravitational potential and, consequently, the coupling parameter $\delta$:
\begin{equation}
    H_{\text{int, finite}} = \frac{m\Omega^2}{2}q^2 - \Lambda \frac{Gm^2}{L+q}
\end{equation}
This leads to a corrected coupling parameter:
\begin{equation}
    \delta = \Lambda \frac{4Gm}{\Omega^2 L^3}.
    \label{eq:delta_lambda}
\end{equation}

The form factor $\Lambda$ represents the ratio of the gravitational force gradient for finite-sized cylinders to that for point masses.
More precisely, it is defined as the ratio of the second derivatives of the gravitational potential energy with respect to the center-of-mass separation $L$: 
\begin{equation}
    \Lambda = \frac{\frac{\partial^2 U_{\text{cylinder}}}{\partial L^2}}{\frac{\partial^2 U_{\text{point-mass}}}{\partial L^2}} = \frac{\frac{\partial^2 U_{\text{cylinder}}}{\partial L^2}}{\frac{2Gm^2}{L^3}}
\end{equation}
where $U_{\text{cylinder}}$ is the gravitational potential energy between two identical, coaxial cylinders, each with mass $m$, radius $R$, and height $h$. Their centers of mass are separated by a distance $L$, resulting in a face-to-face separation of $d=L-h$. 

The calculation of $U_{\text{cylinder}}$ involves integrating the Newtonian potential over the volumes of both cylinders. While the full derivation is extensive, the resulting expression for $\Lambda$ can be formulated as a function of two dimensionless geometric ratios: the radius-to-height ratio ($R/h$) and the center-of-mass–to-height ratio ($L/h$). 

Based on established methods for calculating gravitational interactions between cylindrical bodies, the form factor $\Lambda$ can be expressed by the following integral:
\begin{equation}
    \Lambda = 8 \frac{(d+h)^3}{R^2 h} \int_0^\infty \left[ \frac{J_1\left(\frac{R}{h}s\right)}{s} \right]^2 e^{-\frac{d}{h}s} \sinh^2\left(\frac{s}{2}\right) ds
    \label{eq:lambda_integral}
\end{equation}
Here, $J_1(x)$ is the Bessel function of the first kind of order one.

%\section{Numerical Results}

\begin{figure}[h!]
    \centering
    \includegraphics[width=0.4\textwidth]{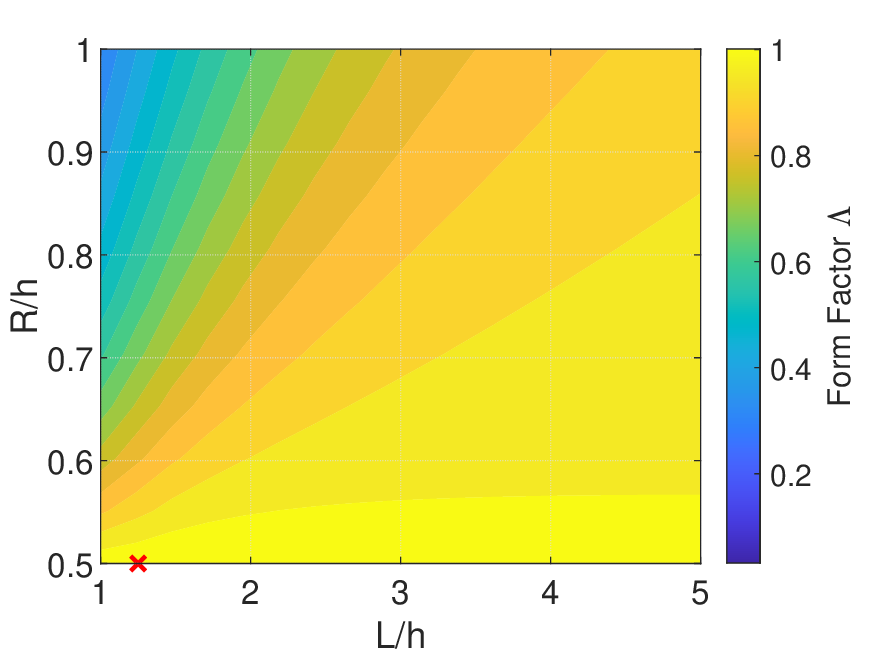}
    \caption{The calculated form factor $\Lambda$ as a function of the dimensionless geometric parameters $L/h$ and $R/h$. The value of $\Lambda$ converges to 1 for large separations, corresponding to the point-mass approximation. The red cross mark corresponds to the parameters selected in this study.}
    \label{fig:form}
\end{figure}

Due to the complexity of the integrand in Eq. \eqref{eq:lambda_integral}, we calculated the form factor $\Lambda$ numerically. The results of this numerical calculation are shown in Figure \ref{fig:form}. The plot illustrates how $\Lambda$ varies as a function of the cylinder geometry. As expected, $\Lambda$ approaches unity for large separations ($L/h \gg 1$), where the finite-size effects become negligible and the cylinders behave like point masses. For closer separations and larger radii, $\Lambda$ deviates significantly from 1, highlighting the necessity of this correction for an accurate analysis of the system.

\section{Covariance matrix}
\label{app:C}
\renewcommand{\theequation}{C\arabic{equation}}

The covariance matrix for individual test masses is extracted from the conditional covariances $V_\pm$ as follows. Among the components of $V_\pm$ representing correlations of $(q_\pm,p_\pm,r_\pm)$, those involving $r_\pm$ are redundant, and the information of the two modes consists essentially of the following.
\begin{align}
\bm{\tilde{V}}_\pm=\left(\begin{array}{cc}
V^\pm_{q q} & V^\pm_{q p} \\
V^\pm_{q p} & V^\pm_{pp}\end{array}\right).
\end{align}
Then, the covariance of the total system $\bm{\mathcal{V}}$, based on the two entangled test masses, is given by~\cite{PhysRevD.110.024057}
%$\bm{\mathcal{V}}$ is given by the formula from the covariance matrix for the common mode ($\bm{V}_+$) and the differential mode ($\bm{V}_-$) by~\cite{PhysRevD.110.024057} 
\begin{eqnarray}
 &&   \bm{\mathcal{V}}\equiv
    \Bigl(\begin{array}{cc}
\bm{\mathcal{V}}_1&\bm{\mathcal{V}}_{12}\\
        \bm{\mathcal{V}}_{12}&\bm{\mathcal{V}}_{2}
    \end{array}\Bigr)
    =\bm{S}
    \Bigl(\begin{array}{cc}
        \bm{\tilde{V}}_+&0
        \label{bmV}
        \\
        0&\bm{\tilde{V}}_{-}
    \end{array}\Bigr)
    \bm{S}^{\text{T}},~~\\
&&    \bm{S}=\frac{1}{\sqrt{2}}
    \left(\begin{array}{cccc}
            1&0&1/(1-\delta)^{1/4}&0\\
        0&1&0&(1-\delta)^{1/4}\\
        1&0&-1/(1-\delta)^{1/4}&0\\
        0&1&0&-(1-\delta)^{1/4}
    \end{array}\right),
\end{eqnarray} 
where $\bm{\mathcal{V}}_1$, $\bm{\mathcal{V}}_2$
are the covariance matrices of test masses 1 and 2 normalized by frequency $\Omega$, respectively. $\bm{\mathcal{V}}_{12}$ represents the gravity-induced correlation matrix between the test masses and $\bm{S}$ is the operation of the beam splitter. 
%Note that $\bm{\mathcal{V}}$ are proportional to the covariance of physical variables. The factor $(1-\delta)^{\pm1/4}$ is required to compensate the difference of the zero point fluctuation between the common and differential modes.  

\section{Cosmic-ray collisions}
\label{app:D}
\renewcommand{\theequation}{D\arabic{equation}}
To evaluate cosmic-ray–induced excitations of the test mass, we perform Monte Carlo simulations using Geant4-11.3.1 \cite{G4} and Root 6.34.08 \cite{G4}.
The test mass is modeled as a cylindrical Pt–Au alloy (27\% Pt and 73\% Au by mass) with a total mass of 1 kg.
The required amplitude spectral density (ASD) of the acceleration noise is calculated using the fluctuation–dissipation theorem.~\cite{PhysRev.83.34}:
\begin{align}
\sqrt{S_a^{(\rm req)}}
&\simeq 
\frac{5.9\times10^{-21}\ {\rm m}}{\rm{s}^2\sqrt{\text{Hz}}}\sqrt{\left(\frac{1~\text{kg}}{\rm{m}}\right)\left(\frac{\Gamma T/2\pi}{2\times10^{-19}~\text{Hz}\cdot  \text{K}}\right)}.
\label{eq:req}
\end{align}

In the simulation, a test mass is placed at the center of the sphere. Protons are injected from the spherical surface with directions cosine-biased relative to the local inward normal. 
%isotropically 
Their energies follow the known galactic cosmic-ray flux distribution \cite{cosmic-ray-flux} \cite{LPF-flux}, with energies ranging from \(10^2\) to \(10^5\) MeV.  
%The shielding structure is designed to decelerate and attenuate primary high-energy protons, and to slow down, thermalize, and absorb secondary particles, primarily neutrons, produced within the shield.  
%The shield consists of concentric spherical layers, arranged in order from the outside: 3 cm of tungsten, 10 cm of boron-loaded polyethylene (\ce{B4C}-PE), 10 cm of pure polyethylene (PE), and 20 cm of iron. 

\begin{figure}
\centering
\includegraphics[width=3.4in]{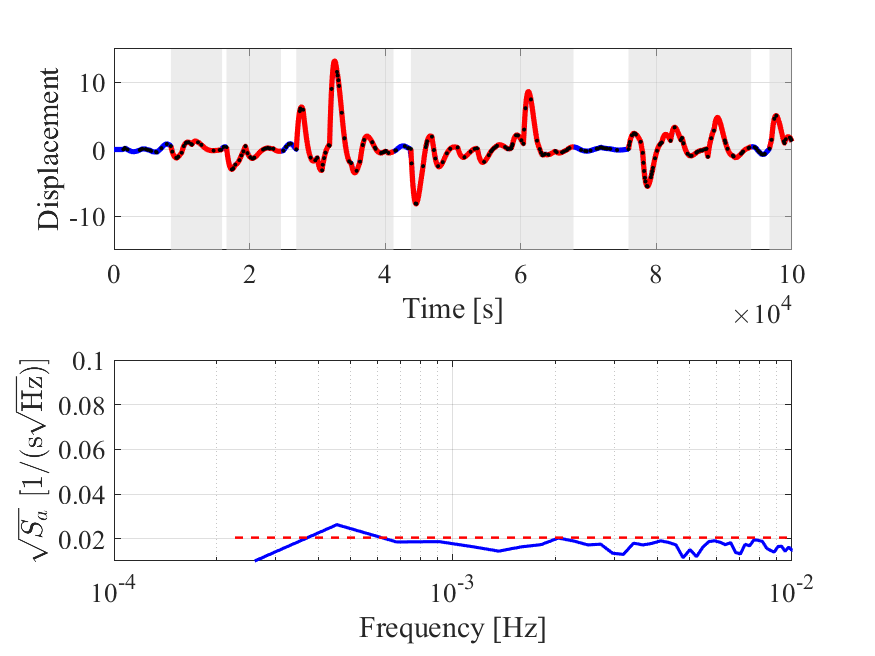}
 \caption{
Simulation of test mass vibrations caused by cosmic-ray collisions, for a mass of 1 kg and $\delta=0.91$. 
Top panel: The time series data for the first 100,000 seconds of the 1.1-million-second simulation. The red segments (on a gray background) indicate regions excluded by data processing. The black dots indicate collision events with cosmic rays.
Bottom panel: The averaged power spectral density obtained exclusively from the blue segments that were not excluded during data processing and whose durations exceed $2\pi/(\delta\Omega)$. It is comparable to the thermal noise level (dashed red). }
\label{fig:Sa}
\end{figure}

We record the momentum kicks delivered to the test mass by cosmic‐ray collisions. To realize GIE, the number of phonons generated by collisions must remain below at most one, throughout the GIE generation time \(2\pi/(\delta\Omega)=4400\)~s for $\Omega/(2\pi)=0.25$~mHz and $\delta=0.91$. 
In practice, if a phonon excitation does occur, measurements are paused until the test mass has returned to its initial state. By increasing the feedback cooling gain \(g_0\) to unity after each excitation, the effective quality factor \(Q\) is reduced to 1. Consequently, the relaxation time scale to the initial state is approximately \(2\pi/\Omega\), and the initialization time is weighted according to the noise magnitude. 
By performing the simulation described above (see the top panel of Fig.~\ref{fig:Sa}), we find that the requirement is satisfied with a duty cycle of 12\%, defined as the fraction of time the measurement process is active within a given period (see the bottom panel of Fig.~\ref{fig:Sa}).
The average total number of large events during a measurement period of $2\pi / (\delta \Omega)$ is 1.9.

\section{Probability density functions}
\label{app:F}
\renewcommand{\theequation}{F\arabic{equation}}

\begin{figure}
\centering
%{\includegraphics[clip, width=3.5in]{hist1_new.eps}}
%{\includegraphics[clip, width=3.5in]{hist2_new.eps}}
{\includegraphics[clip, width=3.5in]{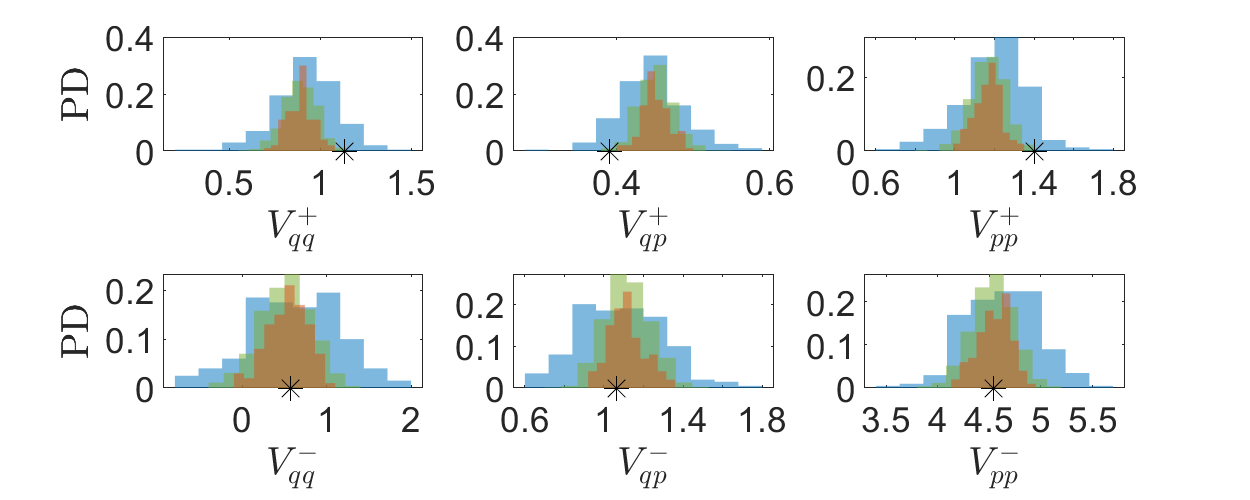}}
{\includegraphics[clip, width=3.5in]{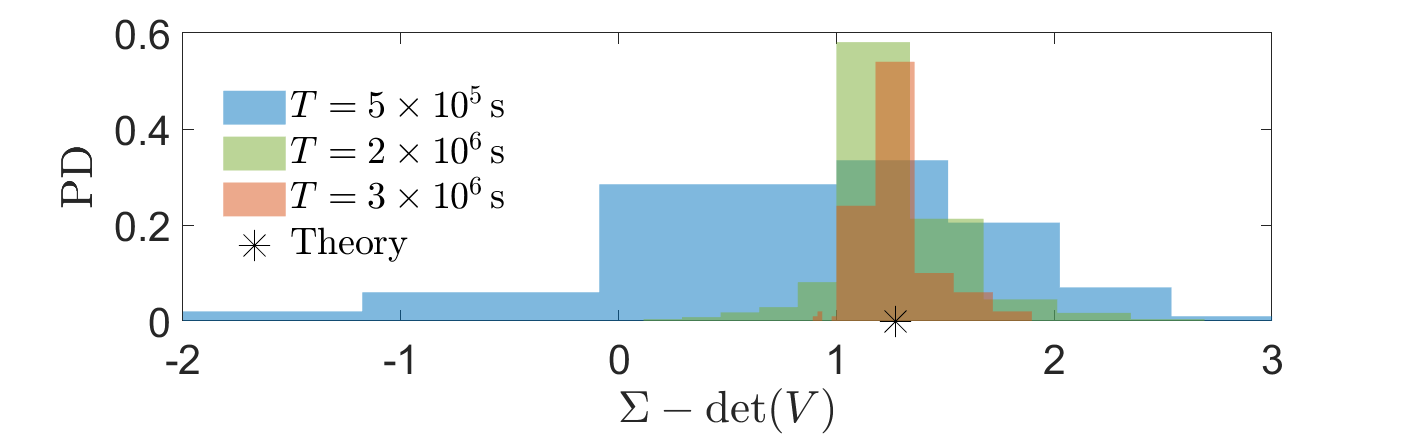}}
{\includegraphics[clip, width=3.5in]{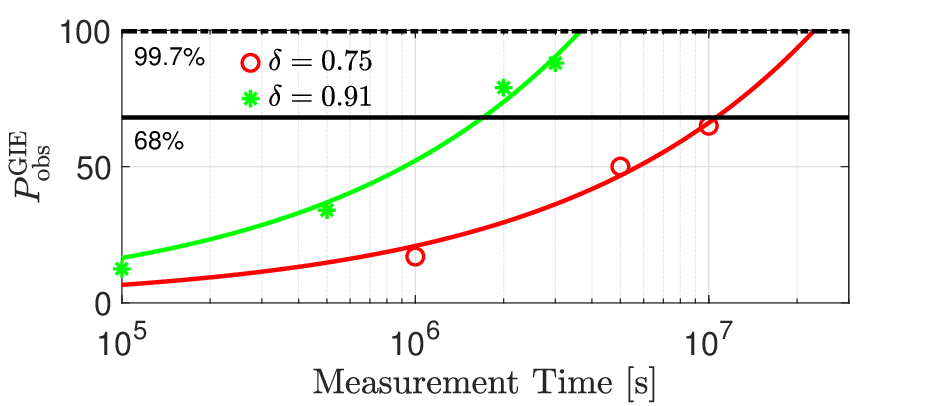}}
\caption{The upper and middle panels display the statistical distributions of $\Sigma - \det(\bm{\mathcal{V}})$, $\bm{V}_{+}$, and $\bm{V}_{-}$. The lower shows $P_{\mathrm{obs}}^{\mathrm{GIE}}$ as a function of measurement time.}
\label{fig:histogram}
\end{figure}

We compute the probability density functions of \( \Sigma - \det(\bm{\mathcal{V}}) \), \( \bm{V}_+ \), and \( \bm{V}_- \) for two distances: $L=1.25h\ (\delta=0.91)$ and $L=4h/3\ (\delta=0.75)$. 
In the top and middle panels of Fig.~\ref{fig:histogram}, we show the probability density functions of \(\Sigma - \det(\bm{\mathcal{V}})\) and of \(\bm{V}_+\) and \(\bm{V}_-\), respectively, for three integration times: \(5\times10^5\)~s (blue), \(2\times10^6\)~s (green), and \(3\times10^6\)~s (red), fot the $\delta=0.91$ case. The asterisks indicate the respective theoretical values.

Compared to the differential mode, the conditional variance of the common mode exhibits a slight deviation from the theoretical value; this asymmetry arises because the laser power is equal in both modes, and the higher resonance frequency of the common mode makes it more susceptible to the influence of shot noise.
In the lower panel of Fig.~\ref{fig:histogram}, circle and star symbols represent \(P_{\mathrm{obs}}^{\mathrm{GIE}}\) for $\delta=0.75$ and $\delta=0.91$, respectively. 
The blue and red lines represent the cases where the cumulative probability increases in proportion to the square root of the integration time.

\section{Passive Shielding for a Common-Mode Inertial Reference}
\label{app:G}

\renewcommand{\theequation}{F\arabic{equation}}

\subsection*{F-1. The Challenge of Common-Mode Acceleration Noise}

As discussed in the main text, our proposed experiment requires simultaneous measurements of both the differential and common-mode motions of two test masses (TMs). While the differential measurement is largely insensitive to the spacecraft’s motion, the common-mode measurement is fundamentally limited by the inertial stability of the spacecraft itself. Consequently, any acceleration of the Science Module (SM) that houses the TMs manifests as common-mode acceleration noise.

For a conventional drag-free satellite such as LPF, the spacecraft motion is dominated by thruster noise (e.g., the ST7 thruster exhibited noise levels of approximately 100~nN/$\sqrt{\mathrm{Hz}}$~\cite{PhysRevD.98.102005}) and by fluctuations in external forces. Because the TMs (with mass $m_{\mathrm{TM}}=1~\mathrm{kg}$) are gravitationally coupled to the SM, any acceleration of the SM, $a_{\mathrm{SM}}$, directly couples to the common mode of the TMs.

To enable GIE detection, the acceleration-noise amplitude spectral density of the 1-kg test masses must satisfy
\begin{equation}
\sqrt{S_{a,\mathrm{TM}}^{(\mathrm{req})}} < 5.9\times10^{-21}~\mathrm{m/s^2/\sqrt{Hz}},
\end{equation}
evaluated at the resonance $f=0.25~\mathrm{mHz}$.
This level is several orders of magnitude below what conventional active drag-free control can achieve, motivating a passively shielded configuration.

\subsection*{F-2. Shielded Formation Concept}

Two modules fly in loose formation:

\begin{enumerate}
\item \textbf{Science Module (SM):}
The 700-kg spacecraft contains the test masses (TMs), auxiliary masses, interferometer, vacuum systems, and refrigeration systems. 
Reference cross-sectional area $A_{\mathrm{SM}}=3.5~\mathrm{m^2}$, giving an equivalent diameter $D_{\mathrm{SM}}=2.11~\mathrm{m}$.

\item \textbf{Shield Module (ShM):}
A tungsten-alloy disk positioned between the Sun and the SM at separation $L=275~\mathrm{m}$.
To maintain a full umbra over the SM,
\begin{equation}
d_{\mathrm{ShM}}\ge D_{\mathrm{SM}}+L\,\theta_\odot,\ \theta_\odot=9.3\times10^{-3}~\mathrm{rad}.
\end{equation}
Hence $d_{\mathrm{ShM}}=4.67~\mathrm{m}$ ensures complete shadowing.
For thickness $t_{\mathrm{ShM}}=0.02~\mathrm{m}$ and density
$\rho=19{,}300~\mathrm{kg/m^3}$,
\begin{equation}
m_{\mathrm{ShM}}=\pi(d/2)^2 t \rho\approx6.6\times10^3~\mathrm{kg}.
\end{equation}
The Sun-facing surface, with a solar absorptance of $\simeq0.1$ and an emissivity of $\simeq0.9$, is passively maintained at approximately $T_{\mathrm{ShM}}\approx227~\mathrm{K}$.
\end{enumerate}

\subsection*{F-2. Transfer Function Between SM and TM}

The apparent common-mode acceleration of each TM, referred to the optical bench on the SM, follows the SM acceleration through the TM–SM coupling.
Modeling this coupling as a lightly damped harmonic link with natural frequency $\Omega$ (set by auxiliary masses) and damping ratio $\zeta=1/(2Q)$ gives
\begin{equation}
H(f)=
\frac{(2\pi f)^2}{(2\pi f)^2-\Omega^2+i\,2\zeta\,\Omega(2\pi f)}.
\label{eq:H_general}
\end{equation}
At the resonance frequency $f_0=\Omega/2\pi$,
the magnitude increases to $|H(f_0)|=1/(2\zeta)=Q$.
This transfer function multiplies SM-originated acceleration noises when referred to the TM readout.
The quality factor $Q$ thus determines how strongly SM motion is transmitted to the apparent TM acceleration.

\subsection*{F-3. Residual Acceleration Noise (SM Level)}

The residual acceleration noise acting on the SM arises from two mechanisms:
(i) gravitational coupling to ShM motion driven by solar-radiation-pressure (SRP) fluctuations, and
(ii) fluctuating thermal radiation pressure.
We adopt $\sqrt{S_T}=0.1~\mathrm{K/\sqrt{Hz}}$
and $\mathcal{G}_{\mathrm{sup}}=10^{-6}$ for geometric suppression.

\subsubsection*{1. Gravitational Coupling from SRP Fluctuations}

The time-varying Newtonian field produced by fluctuations of the Shield Module (ShM)
acts on both the Science Module (SM) and the TMs.
Because the ShM--SM separation ($L$) is hundreds of meters whereas the internal TM--SM spacing ($s$) is only on the order of tens of centimeters,
the gravitational acceleration from the ShM is almost identical at the two locations.
As a result, the SM and TMs experience nearly the same instantaneous acceleration,
so that in the relative (TM--SM) coordinate this common term cancels to first order.
The residual signal arises only from the \emph{gradient} of the ShM’s field,
that is, from the small difference in gravitational pull across the TM--SM baseline.

Expanding $g(r)=Gm_{\rm Sh}/r^2$ about $r=L$ and retaining the leading gradient term
$({\rm d}g/{\rm d}r)_{L}\,s = -2Gm_{\rm Sh}s/L^3$,
a small displacement $x_{\rm Sh}$ of the ShM produces a differential acceleration between the TM and SM of amplitude
$\delta a_{\rm rel} \simeq 6Gm_{\rm Sh}s\,x_{\rm Sh}/L^4$.
Relating the ShM displacement noise to its force noise through
$x_{\rm Sh}=\sqrt{S_{F,{\rm ShM}}}/(m_{\rm Sh}\omega^2)$ then gives
\begin{equation}
\sqrt{S_{a,{\rm TM}\!-\!{\rm SM}}^{\rm (ShM\ grav.)}}
\simeq \frac{6\,G\,s}{L^4}\,
\frac{\sqrt{S_{F,{\rm ShM}}}}{\omega^2},\qquad
\omega=2\pi f.
\end{equation}
Numerically this yields
$\sim 2\times10^{-23}~\mathrm{m/s^2/\sqrt{Hz}}$
for $f=0.25~\mathrm{mHz}$,
$L=275~\mathrm{m}$,
$s\simeq0.5~\mathrm{m}$,
and $\sqrt{S_{F,{\rm ShM}}}=1.7~\mathrm{nN/\sqrt{Hz}}$.
This coupling path does \emph{not} include the $H(f)$ gain factor,
because the first-order common acceleration of the SM and TM is suppressed
by common-mode cancellation, leaving only the much smaller tidal (gravity-gradient) term in the relative readout.

\subsubsection*{2. Thermal Radiation Pressure}

Temperature fluctuations on the Sun-facing surface of the Shield Module
lead to variations in its thermal emission, producing a fluctuating radiation
pressure on the Science Module.
This effect is strongly suppressed by geometry and baffling, but its residual
contribution can be estimated by linearizing the Stefan–Boltzmann law
with respect to temperature fluctuations $\delta T$.
The corresponding acceleration noise on the SM is given by
\begin{align}
\sqrt{S_{a,\mathrm{SM}}^{\mathrm{th}}}
  &= 
  \frac{1}{m_{\mathrm{SM}}}
  \left(
    \frac{4\,\varepsilon\,\sigma\,T_{\mathrm{ShM}}^{3}}{c}
  \right)
  \left(
    \frac{A_{\mathrm{ShM}}}{4\pi L^{2}}
  \right)
  \sqrt{S_{T}}\,
  G_{\mathrm{sup}}\notag\\
&\approx7\times10^{-23}~\mathrm{m/s^2/\sqrt{Hz}}.
\end{align}
This SM-only path \emph{does} map to the TM readout with the transfer gain $|H|$.

This expression assumes isotropic thermal emission from the shield surface,
with $\mathcal{G}_{\mathrm{sup}}$ representing the net geometric suppression
factor due to multi-stage baffling and low-emissivity internal coatings.

\subsection*{F-4. Summary and Optimum $Q$}

The TM-side total at $f=0.25~\mathrm{mHz}$ combines
(i) the gravity-gradient residue from the ShM (no $H$ factor) and
(ii) the SM-only thermal path scaled by $|H|=Q$:
\begin{equation}
\sqrt{S_{a,\mathrm{TM}}^{\mathrm{total}}}
=\sqrt{\left(\sqrt{S_{a,{\rm TM}\!-\!{\rm SM}}^{\rm (ShM\ grav.)}}\right)^2
+\left(Q\,\sqrt{S_{a,\mathrm{SM}}^{\mathrm{th}}}\right)^2}\,.
\end{equation}
With the present numbers
$\sqrt{S_{a,{\rm TM}\!-\!{\rm SM}}}^{\rm (ShM\ grav.)}\!\approx2\times10^{-23}$ and
$\sqrt{S_{a,\mathrm{SM}}^{\mathrm{th}}}\!\approx7\times10^{-23}$,
the requirement $\sqrt{S_{a,\mathrm{TM}}^{(\mathrm{req})}}=5.9\times10^{-21}$ is satisfied provided
\begin{equation}
Q \ \le\ \frac{\sqrt{S_{a,\mathrm{TM}}^{(\mathrm{req})}}}
{\sqrt{S_{a,\mathrm{SM}}^{\mathrm{th}}}}
\ \approx\ \frac{5.9\times10^{-21}}{7\times10^{-23}}
\ \approx\ 8.4\times10^{1}\,.
\end{equation}
Thus, the design easily meets the requirement for any practical $Q$ up to $\mathcal{O}(10^2)$ under the present thermal assumptions.
If, for conservatism, one ignores common-mode cancellation and upper-bounds the SM-only path by the absolute SM acceleration level $\sim4.4\times10^{-21}~\mathrm{m/s^2/\sqrt{Hz}}$, the resulting limit is $Q\lesssim1.3$, which remains achievable.

The present configuration — a 4.7 m-diameter, 6.6 t tungsten shield located at L=275 m — therefore satisfies the common-mode inertial-stability requirement. 
Future work will further refine the design by (i) quantifying the geometric and control tolerances needed to maintain full umbra and formation stability, (ii) developing a detailed force-noise and thermal-fluctuation budget for both the shield and the spacecraft, and (iii) assessing the feasibility of power transfer and attitude control under realistic operational conditions.
\end{document}